\documentclass[preprint,nopreprintnumbers,titlepage,amsmath,amssymb,aps,pre]{revtex4-2} 

\usepackage{amssymb}
\usepackage{amsmath}
\usepackage{txfonts}
\usepackage{mathdots}
\usepackage[classicReIm]{kpfonts}
\usepackage{graphicx}

\newcommand{\expect}[1]{\operatorname{E}\left[#1\right]}   

\newcommand{\Ccal}{\mathcal{C}}

\newcommand{\Hcal}{\mathcal{H}}

\newcommand{\Ncal}{\mathcal{N}}

\newcommand{\Nsc}{N_{\rm sc}}
\newcommand{\Nfft}{N_{\rm fft}}
\newcommand{\Ncp}{N_{\rm cp}}

\newcommand{\jrm}{{\rm j}}

\newcommand{\rect}{{\rm{rect}}}
\newcommand{\SINR}{{\rm{SINR}}}
\newcommand{\SNR}{{\rm{SNR}}}

\newtheorem{Lem}{Lemma}

\begin{document}

\title{A note on simplified SINR expressions for OFDM with insufficient CP}
\author{Renaud-Alexandre Pitaval}
\email{renaud.alexandre.pitaval@huawei.com}
\affiliation{Huawei Technologies Sweden AB\\ SE-164 94 Kista, Sweden}

\begin{abstract}
This note provides derivation details of simplified OFDM transmission equation and resulting signal-to-interference plus noise ratio (SINR) for the case of an insufficient CP. Each channel component after demodulation is expressed as a single sum which can be interpreted a weighted Fourier transform of the channel impulse response. 
Specifically, for CP length $\Ncp$, FFT size $\Nfft$, $\Nsc$ consecutively allocated subcarriers, and channel impulse response $h[n]=\sum^{L_u}_{l=-L_d}{h_l}{\delta }_{n-l}\ $ with $0\le L_d, L_u \le \Nfft-1$; 
the SINR at the $i$th subcarrier is 
\[\SINR_i=\frac{|\Hcal_{0,l,i}|^2 }
{\sum^{\Nsc-1}_{\substack{l=0\\l\neq i}}|\Hcal_{0,l,i}|^2+\sum^{\Nsc-1}_{l=0}|\tilde{\Hcal}_{-1,l,i}|^2+|\tilde{\Hcal}_{1,l,i}|^2+1/\mathrm{SNR}}
\] 
where
\[\Hcal_{0,i,i}=\sum^{L_u}_{m=-L_d}{c[m]h[m]e^{-j2\pi \frac{im}{\Nfft}}} 
\quad\quad
\Hcal_{0,l,i}=\ \sum^{L_u}_{m=-L_d}{{\tilde{c}}_{l,i}[m]h[m]e^{-j2\pi \frac{lm}{\Nfft}}}\] 

\[\tilde{\Hcal}_{-1,i,i}=\sum^{L_u}_{m=\Ncp}{(1-\ c[m]\ )h[m]e^{-j2\pi \frac{im}{\Nfft}}}
\quad\quad
\tilde{\Hcal}_{-1,l,i}=\sum^{L_u}_{m=\Ncp}{{\tilde{c}}_{l,i}[m]h[m]}e^{-j2\pi \frac{lm}{\Nfft}} \] 

\[\tilde{\Hcal}_{1,i,i}=\sum^0_{m={-L}_d}{(1-\ c[m]\ )h[m]}e^{-j2\pi \frac{im}{\Nfft}}
\quad\quad
\tilde{\Hcal}_{1,l,i}=\sum^0_{m={-L}_d}{{\tilde{c}}_{l,i}[m]h[m]}e^{-j2\pi \frac{lm}{\Nfft}} \] 
with weight functions
 \[c[m]=
 \begin{cases} \frac{\Nfft+m}{\Nfft} & -\Nfft \le m\le 0 \\ 
1 & 0\le m\le \Ncp \\ 
\frac{\Nfft -(m-\Ncp)}{\Nfft} & \Ncp\le m\le \Ncp+\Nfft \end{cases}
\quad
\tilde{c}_{l,i}[m]=  \begin{cases}
\frac{1-e^{j2\pi \frac{m(l-i)}{\Nfft}}}{\Nfft(1- e^{j2\pi \frac{(l-i)}{\Nfft}})} & -\Nfft\le m\le 0 \\ 
0 & 0\le m\le \Ncp \\ 
\frac{ e^{j2\pi \frac{(m-\Ncp)(l-i)}{\Nfft}}-1}{\Nfft(1-e^{j2\pi \frac{(l-i)}{\Nfft}})} & \Ncp\le m\le \Nfft+\Ncp \end{cases}
.\]

Moreover, if the channel is causal, i.e.,  $L_d =0$, the SINR depends only on ICI terms as 
\[
\SINR_i = \frac{ |\Hcal_{0,i,i}|^2}
{|\Hcal_i - \Hcal_{0,i,i} |^2 +\sum_{\substack{l=0\\l\neq i}}^{\Nsc-1} 2 |\Hcal_{0,l,i}|^2
												  + 1/\SNR}
\] 
where $
\Hcal_i = \sum^{L_u}_{m=0}{h[m]}e^{-j2\pi \frac{im}{\Nfft}} $
is the Fourier transform of the channel at the $i$th subcarrier.

\end{abstract}

\maketitle

Simplifications of OFDM channels can be found, but scattered, in classical 90's OFDM literature~\cite{Viterbo96,Seoane97,Steendam99}. Still, these derivation steps are often reproduced in recent works for specific scenarios, see e.g.~\cite{aminjavaheri2017ofdm,PitavalGlobecom18}. 
In~\cite{PhamPhD}, detailed derivations of OFDM channels as single sum expressions were summarized for the case of a causal channel. Writing the channel coefficients as such directly  greatly simplifies the resulting SINR expression. 

In this note, we review these derivations with some additional simplifications and extensions.
  
\section{ Transmission equations}
 
An OFDM modulation is defined by a subcarrier spacing $\Delta_f$, an IFFT size $N_{\rm fft}$, and a CP length $\Ncp$.  An OFDM symbol (without CP) has then a time duration of $T_{\rm s}= 1/\Delta_f$ with sampling period $T_{\rm sp}=T_{\rm s}/ \Nfft$. We consider that a total of $\Nsc$ consecutive subcarriers are allocated by i.i.d. data symbols with average power $P$. 

\subsection{OFDM transmission}

The OFDM signal is a consecutive transmission of OFDM blocks as 
\begin{equation}
s[k] =  \sum_{b} s_b[k-b(\Ncp +\Nfft)] 
\end{equation}  
where the data symbol $x_{b,l}$ for the $l$th subcarrier of the $b$th OFDM block is modulated as 
\begin{equation} \label{eq:si}
s_b[k] = \frac{1}{\sqrt{\Nsc}} \sum_{l=0}^{\Nsc-1} x_{b,l} e^{j 2 \pi \frac{lk}{\Nfft}} \rect(k)
 \end{equation}  
with OFDM rectangular block window $\rect(n)= \begin{cases}
1, & -\Ncp\le n\le \Nfft -1  \\ 
0, & \mathrm{otherwise} \end{cases}
.$

\subsection{Channel}
The transmitted signal is convolved with the channel $h[n]$ where its channel impulse response (CIR) is $h[n]=\sum^{L_u}_{l=-L_d}{h_l}{\delta }_{n-l}\ $ with  $-L_d\le 0\le L_u$. The channel taps are assumed to be i.i.d. with average energy of the $p$th channel tap being $E_p=\expect{ |h[p]|^2}$.

The time $n=0$ corresponds to the time of reference (TOR) for demodulation, resulting from synchronization. Therefore, the channel as observed by the receiver may not be causal. This can be, for example, the consequence of using non-causal pulse shaping filtering at the transmitter, such as a sinc shaped filter, and synchronizing the receiver with the filter's maximum peak.

\subsection{Received and demodulated signal} 
The received signal is   
\begin{equation}
r[k]  = \sum^{L_u}_{m=-L_d} h[m] s[k-m] +z[k]
\end{equation}
where $z[k]\sim \Ccal \Ncal(0,\sigma_{z}^2)$ is a zero-mean  additive white Gaussian noise (AWGN) with variance $\sigma_{z}^2$.  After CP removal, for $0\leq k \leq (\Nfft-1)$ we have
\begin{equation}
r[k] =\sum_{b} \sum^{L_u}_{m=-L_d}{h[m]s_b[k-m-b(\Nfft+\Ncp)]}
\end{equation}
and then by substitution of~\eqref{eq:si}  
 \begin{equation}
r[k]=
\frac{1}{\sqrt{\Nsc}}\sum_b\sum^{\Nsc-1}_{l=0} x_{b,l}\left( \sum^{L_u}_{m=-L_d} h[m]e^{-j2\pi \frac{lm}{\Nfft}}\rect[k-m-b(\Nfft+\Ncp)]\right) e^{j2\pi \frac{l(k-b\Ncp) }{\Nfft}}  +z[k].
\end{equation}

The received signal is then demodulated by FFT which gives   the demodulated symbol    for the $i$th subcarrier       
\begin{eqnarray*}
y[i] &=& \frac{\sqrt{\Nsc}}{\Nfft} \sum_{k=0}^{\Nfft-1} r[k] e^{-\jrm2 \pi \frac{ik}{\Nfft}} \\
&=& \Hcal_{0,i,i} x_{0,i} + \underbrace{\sum_{\substack{l=0\\l\neq i}}^{\Nsc-1} \Hcal_{0,l,i} x_{0,l}}_\text{ICI} + 
\underbrace{\sum_{\substack{l=0}}^{\Nsc-1} \sum_{b\neq0} \Hcal_{b,l,i} x_{b,l}}_\text{ISI} + n[i]
\end{eqnarray*}
where for any $b$ 
\begin{equation}
\label{eq:Hbli_DS}
\Hcal_{b,l,i} = \frac{1}{\Nfft} \sum^{\Nfft-1}_{k=0}\sum^{L_u}_{m=-L_d}h[m]e^{-j2\pi \frac{lm-k(l-i)+bl\Ncp}{\Nfft}}\rect[k-m-b(\Nfft+\Ncp)],
\end{equation} 
and $n[i]= \frac{\sqrt{\Nsc}}{\Nfft} \sum_{k=0}^{\Nfft-1} z[k] e^{-\jrm2 \pi \frac{ik}{\Nfft}}$ is the post-processed AWGN with variance $\sigma^2_n = \frac{\Nsc}{\Nfft} \sigma_z^2$.

\section{SINR} 

Accordingly, the signal to interference plus noise ratio (SINR) on the $i$th subcarrier is 
\begin{equation}
\label{eq:SINRi}
\SINR_i = \frac{ |\Hcal_{0,i,i}|^2}{\sum_{\substack{l=0\\l\neq i}}^{\Nsc-1}  |\Hcal_{0,l,i}|^2
												+\sum_{l=0}^{\Nsc-1}  \sum_{b\neq0} |\Hcal_{b,l,i}|^2  + 1/\SNR}, 
\end{equation} 
 with $\Hcal_{b,l,i}$ given in~\eqref{eq:Hbli_DS} and $\SNR = P/\sigma^2_n$.

Below, the terms $\Hcal_{b,l,i}$ are simplified.

\subsection{Simplified channel coefficients}

Each channel component after demodulation can be expressed as a single sum, which can be interpreted as a Fourier transform of
a weighted channel impulse response. Precisely, the desired signal channel on the $i$th subcarrier is 
\begin{equation}
\Hcal_{0,i,i}=\sum^{L_u}_{m=-L_d}c[m]h[m]e^{-j2\pi \frac{im}{\Nfft}}, 
\end{equation}
the ICI channel coefficient from the $l\neq i$ subcarrier is
\begin{equation}
\Hcal_{0,l,i}=\sum^{L_u}_{m=-L_d}\tilde{c}_{l,i}[m] h[m]e^{-j2\pi \frac{lm}{\Nfft}} 
\end{equation}
the ISI channel from the $l$th subcarrier of the $b$-block is
 \begin{equation}
\Hcal_{b,l,i}=\sum^{L_u}_{m=-L_d}a_{b,l,i}[m]h[m]e^{-j2\pi \frac{lm}{\Nfft}} 
\end{equation}
with weight functions 
\begin{equation}
c[m]=\begin{cases}
\frac{\Nfft+m}{\Nfft} & -\Nfft \le m\le 0 \\ 
1 & 0\le m\le \Ncp \\ 
\frac{\Nfft -(m-\Ncp)}{\Nfft} & \Ncp\le m\le \Ncp+\Nfft \\
0 & \text{otherwise}\end{cases}, 
\end{equation}

\begin{equation}
\tilde{c}_{l,i}[m]= \begin{cases}
\frac{1-e^{j2\pi \frac{m(l-i)}{\Nfft}}}{\Nfft(1-e^{j2\pi \frac{(l-i)}{\Nfft}})} & -\Nfft\le m\le 0 \\ 
0 & 0\le m\le \Ncp \\ 
\frac{e^{j2\pi \frac{(m-\Ncp)(l-i)}{\Nfft}}-1}{\Nfft(1-e^{j2\pi \frac{(l-i)}{\Nfft}})} & \Ncp\le m\le \Nfft+\Ncp \\
0 & \text{otherwise} \end{cases} ,
\end{equation}
and
 \begin{equation}
a_{b,i,i}[m]= \begin{cases} 
e^{-j2\pi \frac{bi\Ncp}{\Nfft}}c[m+b(\Nfft+\Ncp)] &  l=i \\
e^{-j2\pi \frac{bl\Ncp}{\Nfft}}\ {\tilde{c}}_{l,i}[m+b(\Nfft+\Ncp)]& l\neq i\end{cases}.
\end{equation}


\subsection{Further simplification with $L_d,L_u \leq \Nfft-1$}

If one can assume that $L_d$ and $L_u$  are both less than a symbol length, i.e., $L_d,L_u \leq \Nfft-1$, the interference  inside one block  then only depends of the previous and consecutive blocks. Thus, we can limit the analysis to $b=-1,0$ and $1$. 

The resulting ISI channel coefficients from the previous block are  
\begin{equation}
\Hcal_{-1,l,i}= 
e^{j2\pi \frac{l\Ncp}{\Nfft}} \times
\begin{cases} 
\displaystyle \sum^{L_u}_{m=\Ncp}{h[m]}e^{-j2\pi \frac{im}{\Nfft}} (1- c[m]) &  l=i \\
\displaystyle -\sum^{L_u}_{m=\Ncp}{h[m]}e^{-j2\pi \frac{lm}{\Nfft}} \tilde{c}_{l,i}[m]& l\neq i
\end{cases},
\end{equation} 
the ISI channel coefficients from the consecutive block are 
\begin{equation}
\Hcal_{-1,l,i}= 
e^{j2\pi \frac{l\Ncp}{\Nfft}} \times
\begin{cases} 
\displaystyle \sum^0_{m=-L_d}{h[m]}e^{-j2\pi \frac{im}{\Nfft}} (1-c[m])&  l=i \\
\displaystyle -\sum^0_{m=-L_d}{h[m]}e^{-j2\pi \frac{lm}{\Nfft}}\tilde{c}_{l,i}[m]& l\neq i
\end{cases},
\end{equation}
and 
\begin{equation} \Hcal_{b,l,i} =0 \text{ for } |b|>1 .
\end{equation}

\subsection{Further simplification with $L_d=0$ and $L_u \leq \Nfft-1$}
If the channel is causal, i.e., the ISI comes only from the previous block with $b=-1$. 
The ISI channel coefficients become
\begin{equation} \Hcal_{b,l,i} =0 \text{ for }  b\neq \{-1,0\}
\end{equation}
and
\begin{equation}
\Hcal_{-1,l,i}= 
e^{j2\pi \frac{l\Ncp}{\Nfft}} \times
\begin{cases} 
\Hcal_i - \Hcal_{0,i,i} &  l=i \\
-\Hcal_{0,l,i} & l\neq i
\end{cases}.
\end{equation}
where 
\begin{equation}
\Hcal_i = \sum^{L_u}_{m=0}{h[m]}e^{-j2\pi \frac{im}{\Nfft}} 
\end{equation}
is the Fourier transform of the channel at the $i$th subcarrier.

This leads to an interference term in the SINR expression that only depends on the ICI power as 
\begin{equation}
\SINR_i = \frac{ |\Hcal_{0,i,i}|^2}
{|\Hcal_i - \Hcal_{0,i,i} |^2 +\sum_{\substack{l=0\\l\neq i}}^{\Nsc-1} 2 |\Hcal_{0,l,i}|^2
												  + 1/\SNR}.
\end{equation}

\section{Average-signal to average-interference plus noise ratio (\MakeLowercase{a}S\MakeLowercase{a}INR)}
Sometimes in literature, see e.g.~\cite{Steendam99,batariere2004cyclic,Mostofi06,aminjavaheri2017ofdm}, the SINR is instead defined as the \emph{average-signal-power} to \emph{average-interference-power}-plus-noise ratio (aSaINR). The information-theoretic justification of this quantity is less clear but it provides often a practical and convenient OFDM design parameter.

Here, we consider again $L_u,L_d\leq \Nfft-1$. 
The aSaINR is then
\begin{equation}
\Gamma_i = \frac{ \expect{|\Hcal_{0,i,i}|^2}}
{\sum^{\Nsc-1}_{\substack{l=0\\l\neq i}}\expect{|\Hcal_{0,l,i}|^2}+\sum^{\Nsc-1}_{l=0}\expect{|\tilde{\Hcal}_{-1,l,i}|^2}+\expect{|\tilde{\Hcal}_{1,l,i}|^2}+1/\mathrm{SNR}}. 
\end{equation} 

Given the average channel tap power as $E_m =\expect{|h_m|^2}$, this  simplifies as 
\begin{equation}
\Gamma_i = \frac{  \sum^{L_u}_{m=L_d} c[m]^2 E_m}
{ \sum^{L_u}_{m=L_d}\left( (1-c[m])^2+\sum^{\Nsc-1}_{\substack{l=0\\l\neq i}}2|\tilde{c}_{l,i}[m]|^2\right)E_m+1/\mathrm{SNR}};
\end{equation} 
which, if $\Nsc = \Nfft$,  further simplifies as
 \begin{equation}
\Gamma_i = \frac{  \sum^{L_u}_{m=L_d} c[m]^2 E_m}
{\sum^{L_u}_{m=L_d}(1-c{[m]}^2)E_m +1/\mathrm{SNR}}.
\end{equation}

 \section{Derivations}
\subsection{Simplification of $\Hcal_{0,l,i} $}
\subsubsection{Case $l=i$}
We start with the simplest case which is desired signal channel 
\begin{eqnarray}
\Hcal_{0,i,i}&=&\frac{1}{\Nfft}\sum^{\Nfft-1}_{k=0}\sum^{L_u}_{m=-L_d}{h[m]e^{-j2\pi \frac{im}{\Nfft}}} \rect(k-m)\\
&=&\sum^{L_u}_{m=-L_d}h[m]e^{-j2\pi \frac{im}{\Nfft}} \times \frac{1}{\Nfft}\sum^{\Nfft-1}_{k=0}\rect(k-m) \\
&=&\sum^{L_u}_{m=-L_d}c[m]h[m]e^{-j2\pi \frac{im}{\Nfft}}
\end{eqnarray}
where 
\begin{equation}
c[m]=\frac{1}{\Nfft}\sum^{\Nfft-1}_{k=0}\rect(k-m)
\end{equation}
which can be computed as
\begin{equation}
c[m]=\begin{cases}
\frac{\Nfft+m}{\Nfft} & -\Nfft \le m\le 0 \\ 
1 & 0\le m\le \Ncp \\ 
\frac{\Nfft -(m-\Ncp)}{\Nfft} & \Ncp\le m\le \Ncp+\Nfft \\
0 & \text{otherwise}\end{cases} .
\end{equation}

\subsubsection{Case $l\neq i$}
Now for the ICI channels, we have 

\begin{eqnarray}
\Hcal_{0,l,i}&=&\frac{1}{\Nfft}\sum^{\Nfft-1}_{k=0}\sum^{L_u}_{m=-L_d}h[m]e^{-j2\pi \frac{lm-k(l-i)}{\Nfft}}\rect[k-m] \\ 
&=&\sum^{L_u}_{m=-L_d}h[m]e^{-j2\pi \frac{lm}{\Nfft}}\times \frac{1}{\Nfft}\sum^{\Nfft-1}_{k=0}e^{j2\pi \frac{k(l-i)}{\Nfft}}\rect[k-m]\\ 
&=&\sum^{L_u}_{m=-L_d}\tilde{c}_{l,i}[m] h[m]e^{-j2\pi \frac{lm}{\Nfft}} 
\end{eqnarray}
where 
\begin{eqnarray}
\tilde{c}_{l,i}[m] &=&\frac{1}{\Nfft}\sum^{\Nfft-1}_{k=0}e^{j2\pi \frac{k(l-i)}{\Nfft}}\rect[k-m]\\
  &=&
	\begin{cases}
 \frac{1}{\Nfft} \sum^{\Nfft-1+m}_{k=0}e^{j2\pi \frac{k(l-i)}{\Nfft}} & -\Nfft\le m\le 0 \\ 
 \frac{1}{\Nfft} \sum^{\Nfft-1}_{k=0}e^{j2\pi \frac{k(l-i)}{\Nfft}}& 0\le m\le \Ncp \\ 
 \frac{1}{\Nfft} \sum^{\Nfft-1}_{k=-(\Ncp-m)}e^{j2\pi \frac{k(l-i)}{\Nfft}}& \Ncp\le m\le \Nfft+\Ncp \end{cases}  .
\end{eqnarray}

Using the formula
\begin{equation}
\sum^a_{k=b}r^k=\frac{r^b-r^{a+1}}{1-r}
\end{equation}
we get
\begin{equation}
\tilde{c}_{l,i}[m]= \begin{cases}
\frac{1-e^{j2\pi \frac{m(l-i)}{\Nfft}}}{\Nfft(1-e^{j2\pi \frac{(l-i)}{\Nfft}})} & -\Nfft\le m\le 0 \\ 
0 & 0\le m\le \Ncp \\ 
\frac{e^{j2\pi \frac{(m-\Ncp)(l-i)}{\Nfft}}-1}{\Nfft(1-e^{j2\pi \frac{(l-i)}{\Nfft}})} & \Ncp\le m\le \Nfft+\Ncp \\
0 & \text{otherwise} \end{cases} .
\end{equation}

\subsection{Simplification of $\Hcal_{b,l,i} $, $b \neq 0$}
For the ISI channels we have 
 \begin{eqnarray}
\Hcal_{b,l,i}&=&\sum^{L_u}_{m=-L_d}h[m]e^{-j2\pi \frac{lm}{\Nfft}}  \times 
\frac{e^{j2\pi \frac{l\Ncp}{\Nfft}}}{\Nfft} \sum^{\Nfft-1}_{k=0}{e^{j2\pi \frac{k(l-i)}{\Nfft}}\rect[k-m-b(\Nfft+\Ncp)]}\\
&=&\sum^{L_u}_{m=-L_d}a_{b,l,i}[m]h[m]e^{-j2\pi \frac{lm}{\Nfft}} 
\end{eqnarray}
where 
\begin{eqnarray}
a_{b,l,i}[m]=\frac{e^{-j2\pi \frac{bl\Ncp}{\Nfft}}}{\Nfft}\ \sum^{\Nfft-1}_{k=0}e^{j2\pi \frac{k(l-i)}{\Nfft}}\rect[k-m-b(\Nfft+\Ncp)].
\end{eqnarray}
Two cases can be distinguished. 
\subsubsection{Case $l=i$}
For the ISI resulting from the same subcarrier index, we have 
\begin{eqnarray}
a_{b,i,i}[m]&=&\frac{e^{-j2\pi \frac{bi\Ncp}{\Nfft}}}{\Nfft}\sum^{\Nfft-1}_{k=0}\rect[k-m-b(\Nfft+\Ncp)]\\
&=&e^{-j2\pi \frac{bi\Ncp}{\Nfft}}c[m+b(\Nfft+\Ncp)] 
\end{eqnarray}
where the last equality follows from $c[m]= \frac{1}{\Nfft}\sum^{\Nfft-1}_{k=0}\rect(k-m)$.

Precisely,
\begin{equation}
a_{b,i,i} =e^{-j2\pi \frac{bi\Ncp}{\Nfft}} \begin{cases}
\frac{m+(b+1)\Nfft+b\Ncp}{\Nfft} & -(1+b)\Nfft-b\Ncp \le m\le -b(\Nfft+\Ncp) \\ 
1 & -b(\Nfft+\Ncp)\le m\le -b\Nfft+(1-b)\Ncp \\ 
\frac{ (1-b)(\Nfft+\Ncp)-m}{\Nfft} & -b\Nfft+(1-b)\Ncp \le m\le (1-b)(\Nfft+\Ncp) \\
0 & \text{otherwise}\end{cases} .
\end{equation}

\subsubsection{Case $l\neq i$}
For the ISI resulting from the other subcarrier indices, we get
\begin{eqnarray}
a_{b,l,i}[m]&=&\frac{e^{-j2\pi \frac{bl\Ncp}{\Nfft}}}{\Nfft}\ \sum^{\Nfft-1}_{k=0}{e^{j2\pi \frac{k(l-i)}{\Nfft}}\rect[k-m-b(\Nfft+\Ncp)]} \\
&=& e^{-j2\pi \frac{bl\Ncp}{\Nfft}}\ {\tilde{c}}_{l,i}[m+b(\Nfft+\Ncp)]
\end{eqnarray}
which is precisely given as 
\begin{equation}
a_{b,l,i} =e^{-j2\pi \frac{bl\Ncp}{\Nfft}} \begin{cases}
\frac{1-e^{j2\pi \frac{(m+b\Ncp)(l-i)}{\Nfft}}}{\Nfft(1-e^{j2\pi \frac{(l-i)}{\Nfft}})} & -(1+b)\Nfft-b\Ncp \le m\le -b(\Nfft+\Ncp) \\ 
0 & -b(\Nfft+\Ncp)\le m\le -b\Nfft+(1-b)\Ncp \\ 
\frac{e^{j2\pi \frac{(m+(b-1)\Ncp)(l-i)}{\Nfft}}-1}{\Nfft(1-e^{j2\pi \frac{(l-i)}{\Nfft}})} & -b\Nfft+(1-b)\Ncp \le m\le (1-b)(\Nfft+\Ncp) \\
0 & \text{otherwise}\end{cases} .
\end{equation}

\subsection{Special cases $L_u,L_d\leq \Nfft-1$ }

Remark that 
\begin{equation}
a_{-1,i,i}=e^{j2\pi \frac{i\Ncp}{\Nfft}}(1-c[m]) \quad \mathrm{for} \quad 0\le m\le \Nfft+2\Ncp,
\end{equation}
\begin{equation}
a_{1,i,i}=e^{j2\pi \frac{i\Ncp}{\Nfft}}(1-c[m]) \quad \mathrm{for} \quad  - \Nfft\le m\le 0 ,
\end{equation}
\begin{equation}
a_{-1,l,i}[m]=-e^{j2\pi \frac{l\Ncp}{\Nfft}}\tilde{c}_{l,i}[m] \quad \mathrm{for} \quad 0\le m\le \Nfft+2\Ncp,
\end{equation}
\begin{equation}
a_{1,l,i}[m]=-\ e^{-j2\pi \frac{l\Ncp}{\Nfft}}{\tilde{c}}_{l,i}[m]  \quad \mathrm{for} \quad - \Nfft\le m\le 0.
\end{equation}

Thus with $L_u,L_d\leq \Nfft-1$, we have as special cases 
\begin{equation} a_{-1,i,i}[m]= \begin{cases}
0 &  -\Nfft\le m\le 0 \\ 
e^{j2\pi \frac{i\Ncp}{\Nfft}}(1- c[m])&  0\le m\le \Nfft 
\end{cases},
\end{equation} 

\begin{equation} a_{-1,l,i}[m]=\begin{cases}
0 & -\Nfft\le m\le \Ncp \\ 
-e^{j2\pi \frac{l\Ncp}{\Nfft}}\tilde{c}_{l,i}[m] & \Ncp\le m\le \Nfft 
\end{cases};
\end{equation} 
and similarly 
\begin{equation} a_{1,i,i}[m]=\begin{cases}
e^{-j2\pi \frac{i\Ncp}{\Nfft}} (1- c[m]) & \Nfft\le m\le 0 \\ 
0 & 0\le m\le \Nfft \end{cases}
\end{equation}  
\begin{equation} a_{1,l,i}[m]=\begin{cases}
-e^{-j2\pi \frac{l\Ncp}{\Nfft}}\tilde{c}_{l,i}[m]&-\Nfft\le m\le 0 \\ 
0 &0\le m\le \Nfft  
\end{cases}.
\end{equation}

The resulting simplied channel coefficients direcly follows from these.


\subsection{aSaINR derivations}
By direct averaging we have the average desired signal power
\begin{equation} \expect{|\Hcal_{0,i,i}|^2}= \sum^{L_u}_{m=L_d} c[m]^2 E_m.
\end{equation} 

Similarly we can average the ICI terms as
\begin{equation}
\expect{|\Hcal_{0,l,i}|^2}=\sum^{L_u}_{m=L_d}E_m |\tilde{c}_{l,i}[m]|^2,
\end{equation}
 and the ISI terms as 
\begin{equation}
\expect{|\Hcal_{-1,i,i}|^2}= \sum^{L_u}_{m=\Ncp} (1-c[m])^2 E_m,
\quad\quad\quad
\expect{|\Hcal_{-1,l,i}|^2}=\sum^{L_u}_{m=\Ncp} |\tilde{c}_{l,i}[m]|^2 E_m, 
\end{equation}

\begin{equation}
\expect{|\Hcal_{1,i,i}|^2}= \sum^0_{m=L_d}(1-c{[m])}^2 E_m,
\quad\quad\quad
\expect{|\Hcal_{1,l,i}|^2}=\sum^0_{m=L_d} |\tilde{c}_{l,i}[m]|^2E_m.
\end{equation}

Therefore, the total interference power is 
\begin{eqnarray}
\expect{I} &=&\sum^{\Nsc-1}_{\substack{l=0\\l\neq i}}\expect{|\Hcal_{0,l,i}|^2}+\sum^{\Nsc-1}_{l=0}\expect{|\tilde{\Hcal}_{-1,l,i}|^2}+\expect{|\tilde{\Hcal}_{1,l,i}|^2}\\
&=& \sum^{L_u}_{m=L_d} (1-c[m])^2E_m+\sum^{L_u}_{m=L_d} \sum^{\Nsc-1}_{\substack{l=0\\l\neq i}}2 |\tilde{c}_{l,i}[m]|^2 E_m\\
&=&\sum^{L_u}_{m=L_d}\left( (1-c[m])^2+\sum^{\Nsc-1}_{\substack{l=0\\l\neq i}}2|\tilde{c}_{l,i}[m]|^2\right)E_m  .
\end{eqnarray}

In the case, $\Nsc= \Nfft$, we can simplify the total interference term as 
\begin{equation}
\expect{I}=\sum^{L_u}_{m=L_d} (1-c[m]^2)E_m
\end{equation}
which follows from the lemma below. 

\newpage
\begin{Lem}
\begin{equation}
\sum^{\Nfft-1}_{ \substack{l=0\\l\neq i}}|\tilde{c}_{l,i}[m]|^2= c[m]-c[m]^2
\end{equation}
\end{Lem}

\emph{Proof:} 
First, by direct expansion we get
\begin{eqnarray}
|\tilde{c}_{l,i}[m]|^2&=&\frac{1}{\Nfft^2}\left|\sum^{\Nfft-1}_{k=0}e^{j2\pi \frac{k(l-i)}{\Nfft}}\rect[k-m]\right|^2 \\ 
&=&\frac{1}{\Nfft^2}\left(\sum^{\Nfft-1}_{k=0}e^{j2\pi \frac{k(l-i)}{\Nfft}}\rect[k-m]\right)\left(\sum^{\Nfft-1}_{h=0}e^{j2\pi \frac{-h(l-i)}{\Nfft}}\rect[h-m]\right) \\
&=&\frac{1}{\Nfft^2}\sum^{\Nfft-1}_{k=0}\sum^{\Nfft-1}_{h=0}e^{j2\pi \frac{(k-h)(l-i)}{\Nfft}}\rect[k-m]\rect[h-m] 
\end{eqnarray}
From this, it follows 
\begin{eqnarray} 
\sum^{\Nfft-1}_{\substack{l=0\\l\neq i}}|\tilde{c}_{l,i}[m]|^2 &=&
\frac{1}{\Nfft^2}\sum^{\Nfft-1}_{k=0}\sum^{\Nfft-1}_{h=0}\sum^{\Nsc-1}_{\substack{l=0\\l\neq i}}e^{j2\pi \frac{(k-h)(l-i)}{\Nfft}}
\rect[k-m]\rect[h-m] \\
&=& \frac{1}{\Nfft^2}\sum^{\Nfft-1}_{k=0}\sum^{\Nfft-1}_{h=0}\sum^{\Nsc-1}_{\eta =1}e^{j2\pi \frac{(k-h)\eta }{\Nfft}}\rect[k-m]\rect[h-m] 
\end{eqnarray}
where in the last equality we used the change of variable $\eta=l-i$.

Now assuming $\Nsc=\Nfft$, recalling that $c[m]=\frac{1}{\Nfft}\sum^{\Nfft-1}_{k=0}\rect(k-m)$, and remarking that
\begin{equation}
\sum^{\Nsc-1}_{\eta =1}e^{j2\pi \frac{(k-h)\eta }{\Nfft}}=\begin{cases}-1 & (k-h)\neq 0 \\\Nfft-1 & (k-h)= 0 \end{cases}
\end{equation},
we get
 
\begin{eqnarray} 
\sum^{\Nfft-1}_{ \substack{l=0\\l\neq i}}|\tilde{c}_{l,i}[m]|^2 &=&
 \frac{1}{\Nfft^2}\sum^{\Nfft-1}_{k=0}\left((\Nfft-1) \times \rect[k-m]^2
+\sum^{\Nfft-1}_{  \substack{h=0\\h\neq k}}(-1)\times \rect[k-m]\rect[h-m]\right) \nonumber\\
  &=&
\frac{(\Nfft-1)}{\Nfft} c[m]-\frac{1}{\Nfft^2}\sum^{\Nfft-1}_{k=0}\rect[k-m]\sum^{\Nfft-1}_{\substack{h=0\\h\neq k}}\rect[h-m] \\
&=& \frac{(\Nfft-1)}{\Nfft} c[m]-\frac{1}{\Nfft^2}\sum^{\Nfft-1}_{k=0}\rect[k-m]\left(\sum^{\Nfft-1}_{h=0}\rect[h-m]-\rect[k-m]\right)\\
&=& \frac{(\Nfft-1)}{\Nfft} c[m]-\frac{1}{\Nfft^2}\sum^{\Nfft-1}_{k=0}\sum^{\Nfft-1}_{h=0}\rect[k-m]\rect[h-m]\\
&&\mkern160mu+\frac{1}{\Nfft^2}\sum^{\Nfft-1}_{k=0}\rect[k-m]\rect[k-m] \\
&=&  \frac{(\Nfft-1)}{\Nfft} c[m]-c[m]^2+\frac{c[m]}{\Nfft}\\ 
&=& c[m]-c[m]^2 
\end{eqnarray}

\addcontentsline{toc}{chapter}{Bibliography}
\bibliography{mybibfile}

\end{document}